\documentstyle[e-e-ijmpa,twoside]{article}
\input psfig    

\def\fileversion{v1.20}%
\def\filedate{26.1.94}%
\edef\epsfigRestoreAt{\catcode`@=\number\catcode`@\relax}%
\catcode`\@=11\relax
\ifx\undefined\@makeother                
\def\@makeother#1{\catcode`#1=12\relax}  
\fi                                      
\immediate\write16{Document style option `epsfig', \fileversion\space
<\filedate> (edited by SPQR)}%
\newcount\EPS@Height \newcount\EPS@Width \newcount\EPS@xscale
\newcount\EPS@yscale
\def\psfigdriver#1{%
  \bgroup\edef\next{\def\noexpand\tempa{#1}}%
    \uppercase\expandafter{\next}%
    \def\LN{DVITOLN03}%
    \def\DVItoPS{DVITOPS}%
    \def\DVIPS{DVIPS}%
    \def\emTeX{EMTEX}%
    \def\OzTeX{OZTEX}%
    \def\Textures{TEXTURES}%
    \global\chardef\fig@driver=0
    \ifx\tempa\LN
        \global\chardef\fig@driver=0\fi
    \ifx\tempa\DVItoPS
        \global\chardef\fig@driver=1\fi
    \ifx\tempa\DVIPS
        \global\chardef\fig@driver=2\fi
    \ifx\tempa\emTeX
        \global\chardef\fig@driver=3\fi
    \ifx\tempa\OzTeX
        \global\chardef\fig@driver=4\fi
    \ifx\tempa\Textures
        \global\chardef\fig@driver=5\fi
  \egroup
\def\psfig@start{}%
\def\psfig@end{}%
\def\epsfig@gofer{}%
\ifcase\fig@driver
\typeout{WARNING! ****
 no specials for LN03 psfig}%
\or 
\def\psfig@start{}%
\def\psfig@end{\special{dvitops: import \@p@sfilefinal \space
\@p@swidth sp \space \@p@sheight sp \space fill}%
\if@clip \typeout{Clipping not supported}\fi
\if@angle \typeout{Rotating not supported}\fi
}%
\let\epsfig@gofer\psfig@end
\or 
\def\psfig@start{\special{ps::[begin]  \@p@swidth \space \@p@sheight \space%
        \@p@sbbllx \space \@p@sbblly \space%
        \@p@sbburx \space \@p@sbbury \space%
        startTexFig \space }%
        \if@angle
                \special {ps:: \@p@sangle \space rotate \space}
        \fi
        \if@clip
                \if@verbose
                        \typeout{(clipped to BB) }%
                \fi
                \special{ps:: doclip \space }%
        \fi
        \special{ps: plotfile \@p@sfilefinal \space }%
        \special{ps::[end] endTexFig \space }%
}%
\def\psfig@end{}%
\def\epsfig@gofer{\if@clip
                        \if@verbose
                           \typeout{(clipped to BB)}%
                        \fi
                        \epsfclipon
                  \fi
                  \epsfsetgraph{\@p@sfilefinal}%
}%
\or 
\typeout{WARNING. You must have a .bb info file with the Bounding Box
  of the pcx file}%
\def\psfig@start{}%
\def\psfig@end{\typeout{pcx import of \@p@sfilefinal}%
\if@clip \typeout{Clipping not supported}\fi
\if@angle \typeout{Rotating not supported}\fi
\raisebox{\@p@srheight sp}{\special{em: graph \@p@sfilefinal}}}%
\def\epsfig@gofer{}%
\or 
\def\psfig@start{}%
\def\psfig@end{%
\EPS@Width\@p@swidth
\EPS@Height\@p@sheight
\divide\EPS@Width by 65781  
\divide\EPS@Height by 65781
\special{epsf=\@p@sfilefinal
\space
width=\the\EPS@Width
\space
height=\the\EPS@Height
}%
\if@clip \typeout{Clipping not supported}\fi
\if@angle \typeout{Rotating not supported}\fi
}%
\let\epsfig@gofer\psfig@end
\or 
\def\psfig@end{
         \EPS@Width=\@bbw  
         \divide\EPS@Width by 1000
         \EPS@xscale=\@p@swidth \divide \EPS@xscale by \EPS@Width
         \EPS@Height=\@bbh  
         \divide\EPS@Height by 1000
         \EPS@yscale=\@p@sheight \divide \EPS@yscale by\EPS@Height
  \ifnum\EPS@xscale>\EPS@yscale\EPS@xscale=\EPS@yscale\fi
\if@clip
   \if@verbose
      \typeout{(clipped to BB)}%
   \fi
   \epsfclipon
\fi
\special{illustration \@p@sfilefinal\space scaled \the\EPS@xscale}%
}%
\def\psfig@start{}%
\let\epsfig\psfig
\else
\typeout{WARNING. *** unknown  driver - no psfig}%
\fi
}%
\newdimen\ps@dimcent
\ifx\undefined\fbox
\newdimen\fboxrule
\newdimen\fboxsep
\newdimen\ps@tempdima
\newbox\ps@tempboxa
\long\def\fbox#1{\leavevmode\setbox\ps@tempboxa\hbox{#1}\ps@tempdima\fboxrule
    \advance\ps@tempdima \fboxsep \advance\ps@tempdima \dp\ps@tempboxa
   \hbox{\lower \ps@tempdima\hbox
  {\vbox{\hrule height \fboxrule
          \hbox{\vrule width \fboxrule \hskip\fboxsep
          \vbox{\vskip\fboxsep \box\ps@tempboxa\vskip\fboxsep}\hskip
                 \fboxsep\vrule width \fboxrule}%
                 \hrule height \fboxrule}}}}%
\fi
\ifx\@ifundefined\undefined
\long\def\@ifundefined#1#2#3{\expandafter\ifx\csname
  #1\endcsname\relax#2\else#3\fi}%
\fi
\@ifundefined{typeout}%
{\gdef\typeout#1{\immediate\write\sixt@@n{#1}}}%
{\relax}%
\@ifundefined{epsfig}{}{\typeout{EPSFIG --- already loaded} }%
\@ifundefined{epsfbox}{\input epsf}{}%
\ifx\undefined\@latexerr
        \newlinechar`\^^J
        \def\@spaces{\space\space\space\space}%
        \def\@latexerr#1#2{%
        \edef\@tempc{#2}\expandafter\errhelp\expandafter{\@tempc}%
        \typeout{Error. \space see a manual for explanation.^^J
         \space\@spaces\@spaces\@spaces Type \space H <return> \space for
         immediate help.}\errmessage{#1}}%
\fi
\def\@whattodo{You tried to include a PostScript figure which
cannot be found^^JIf you press return to carry on anyway,^^J
The failed name will be printed in place of the figure.^^J
or type X to quit}%
\def\@whattodobb{You tried to include a PostScript figure which
has no^^Jbounding box, and you supplied none.^^J
If you press return to carry on anyway,^^J
The failed name will be printed in place of the figure.^^J
or type X to quit}%
\def\@nnil{\@nil}%
\def\@empty{}%
\def\@psdonoop#1\@@#2#3{}%
\def\@psdo#1:=#2\do#3{\edef\@psdotmp{#2}\ifx\@psdotmp\@empty \else
    \expandafter\@psdoloop#2,\@nil,\@nil\@@#1{#3}\fi}%
\def\@psdoloop#1,#2,#3\@@#4#5{\def#4{#1}\ifx #4\@nnil \else
       #5\def#4{#2}\ifx #4\@nnil \else#5\@ipsdoloop #3\@@#4{#5}\fi\fi}%
\def\@ipsdoloop#1,#2\@@#3#4{\def#3{#1}\ifx #3\@nnil
       \let\@nextwhile=\@psdonoop \else
      #4\relax\let\@nextwhile=\@ipsdoloop\fi\@nextwhile#2\@@#3{#4}}%
\def\@tpsdo#1:=#2\do#3{\xdef\@psdotmp{#2}\ifx\@psdotmp\@empty \else
    \@tpsdoloop#2\@nil\@nil\@@#1{#3}\fi}%
\def\@tpsdoloop#1#2\@@#3#4{\def#3{#1}\ifx #3\@nnil
       \let\@nextwhile=\@psdonoop \else
      #4\relax\let\@nextwhile=\@tpsdoloop\fi\@nextwhile#2\@@#3{#4}}%
\long\def\epsfaux#1#2:#3\\{\ifx#1\epsfpercent
   \def\testit{#2}\ifx\testit\epsfbblit
        \@atendfalse
        \epsf@atend #3 . \\%
        \if@atend
           \if@verbose
                \typeout{epsfig: found `(atend)'; continuing search}%
           \fi
        \else
                \epsfgrab #3 . . . \\%
                \epsffileokfalse\global\no@bbfalse
                \global\epsfbbfoundtrue
        \fi
   \fi\fi}%
\def\epsf@atendlit{(atend)}
\def\epsf@atend #1 #2 #3\\{%
   \def\epsf@tmp{#1}\ifx\epsf@tmp\empty
      \epsf@atend #2 #3 .\\\else
   \ifx\epsf@tmp\epsf@atendlit\@atendtrue\fi\fi}%

\chardef\trig@letter = 11
\chardef\other = 12
 
\newif\ifdebug 
\newif\ifc@mpute 
\newif\if@atend
\c@mputetrue 
 
\let\then = \relax
\def\r@dian{pt }%
\let\r@dians = \r@dian
\let\dimensionless@nit = \r@dian
\let\dimensionless@nits = \dimensionless@nit
\def\internal@nit{sp }%
\let\internal@nits = \internal@nit
\newif\ifstillc@nverging
\def \Mess@ge #1{\ifdebug \then \message {#1} \fi}%
 
{ 
        \catcode `\@ = \trig@letter
        \gdef \nodimen {\expandafter \n@dimen \the \dimen}%
        \gdef \term #1 #2 #3%
               {\edef \t@ {\the #1}
                \edef \t@@ {\expandafter \n@dimen \the #2\r@dian}%
                \t@rm {\t@} {\t@@} {#3}%
               }%
        \gdef \t@rm #1 #2 #3%
               {{%
                \count 0 = 0
                \dimen 0 = 1 \dimensionless@nit
                \dimen 2 = #2\relax
                \Mess@ge {Calculating term #1 of \nodimen 2}%
                \loop
                \ifnum  \count 0 < #1
                \then   \advance \count 0 by 1
                        \Mess@ge {Iteration \the \count 0 \space}%
                        \Multiply \dimen 0 by {\dimen 2}%
                        \Mess@ge {After multiplication, term = \nodimen 0}%
                        \Divide \dimen 0 by {\count 0}%
                        \Mess@ge {After division, term = \nodimen 0}%
                \repeat
                \Mess@ge {Final value for term #1 of
                                \nodimen 2 \space is \nodimen 0}%
                \xdef \Term {#3 = \nodimen 0 \r@dians}%
                \aftergroup \Term
               }}%
        \catcode `\p = \other
        \catcode `\t = \other
        \gdef \n@dimen #1pt{#1} 
}%
 
\def \Divide #1by #2{\divide #1 by #2} 
 
\def \Multiply #1by #2
       {{
        \count 0 = #1\relax
        \count 2 = #2\relax
        \count 4 = 65536
        \Mess@ge {Before scaling, count 0 = \the \count 0 \space and
                        count 2 = \the \count 2}%
        \ifnum  \count 0 > 32767 
        \then   \divide \count 0 by 4
                \divide \count 4 by 4
        \else   \ifnum  \count 0 < -32767
                \then   \divide \count 0 by 4
                        \divide \count 4 by 4
                \else
                \fi
        \fi
        \ifnum  \count 2 > 32767 
        \then   \divide \count 2 by 4
                \divide \count 4 by 4
        \else   \ifnum  \count 2 < -32767
                \then   \divide \count 2 by 4
                        \divide \count 4 by 4
                \else
                \fi
        \fi
        \multiply \count 0 by \count 2
        \divide \count 0 by \count 4
        \xdef \product {#1 = \the \count 0 \internal@nits}%
        \aftergroup \product
       }}%
 
\def\r@duce{\ifdim\dimen0 > 90\r@dian \then   
                \multiply\dimen0 by -1
                \advance\dimen0 by 180\r@dian
                \r@duce
            \else \ifdim\dimen0 < -90\r@dian \then  
                \advance\dimen0 by 360\r@dian
                \r@duce
                \fi
            \fi}%
 
\def\Sine#1%
       {{%
        \dimen 0 = #1 \r@dian
        \r@duce
        \ifdim\dimen0 = -90\r@dian \then
           \dimen4 = -1\r@dian
           \c@mputefalse
        \fi
        \ifdim\dimen0 = 90\r@dian \then
           \dimen4 = 1\r@dian
           \c@mputefalse
        \fi
        \ifdim\dimen0 = 0\r@dian \then
           \dimen4 = 0\r@dian
           \c@mputefalse
        \fi
        \ifc@mpute \then
                \divide\dimen0 by 180
                \dimen0=3.141592654\dimen0
                \dimen 2 = 3.1415926535897963\r@dian 
                \divide\dimen 2 by 2 
                \Mess@ge {Sin: calculating Sin of \nodimen 0}%
                \count 0 = 1 
                \dimen 2 = 1 \r@dian 
                \dimen 4 = 0 \r@dian 
                \loop
                        \ifnum  \dimen 2 = 0 
                        \then   \stillc@nvergingfalse
                        \else   \stillc@nvergingtrue
                        \fi
                        \ifstillc@nverging 
                        \then   \term {\count 0} {\dimen 0} {\dimen 2}%
                                \advance \count 0 by 2
                                \count 2 = \count 0
                                \divide \count 2 by 2
                                \ifodd  \count 2 
                                \then   \advance \dimen 4 by \dimen 2
                                \else   \advance \dimen 4 by -\dimen 2
                                \fi
                \repeat
        \fi
                        \xdef \sine {\nodimen 4}%
       }}%
 
\def\Cosine#1{\ifx\sine\UnDefined\edef\Savesine{\relax}\else
                             \edef\Savesine{\sine}\fi
        {\dimen0=#1\r@dian\multiply\dimen0 by -1
         \advance\dimen0 by 90\r@dian
         \Sine{\nodimen 0}%
         \xdef\cosine{\sine}%
         \xdef\sine{\Savesine}}}
\def\psdraft{\def\@psdraft{0}}%
\def\psfull{\def\@psdraft{1}}%
\psfull
\newif\if@compress
\def\pscompress{\@compresstrue}
\def\psnocompress{\@compressfalse}
\@compressfalse
\newif\if@scalefirst
\def\psscalefirst{\@scalefirsttrue}%
\def\psrotatefirst{\@scalefirstfalse}%
\psrotatefirst
\newif\if@draftbox
\def\psnodraftbox{\@draftboxfalse}%
\@draftboxtrue
\newif\if@noisy
\@noisyfalse
\newif\ifno@bb
\newif\if@bbllx
\newif\if@bblly
\newif\if@bburx
\newif\if@bbury
\newif\if@height
\newif\if@width
\newif\if@rheight
\newif\if@rwidth
\newif\if@angle
\newif\if@clip
\newif\if@verbose
\newif\if@prologfile
\def\@p@@sprolog#1{\@prologfiletrue\def\@prologfileval{#1}}%
\def\@p@@sclip#1{\@cliptrue}%
\newif\ifepsfig@dos  
\def\epsfigdos{\epsfig@dostrue}%
\epsfig@dosfalse
\newif\ifuse@psfig
\def\ParseName#1{\expandafter\@Parse#1}%
\def\@Parse#1.#2:{\gdef\BaseName{#1}\gdef\FileType{#2}}%

\def\@p@@sfile#1{%
  \ifepsfig@dos
     \ParseName{#1:}%
  \else
     \gdef\BaseName{#1}\gdef\FileType{}%
  \fi
  \def\@p@sfile{NO FILE: #1}%
  \def\@p@sfilefinal{NO FILE: #1}%
  \openin1=#1
  \ifeof1\closein1\openin1=\BaseName.bb
    \ifeof1\closein1
      \if@bbllx                 
        \if@bblly\if@bburx\if@bbury
          \def\@p@sfile{#1}%
          \def\@p@sfilefinal{#1}%
        \fi\fi\fi
      \else                     
        \@latexerr{ERROR. PostScript file #1 not found}\@whattodo
        \@p@@sbbllx{100bp}%
        \@p@@sbblly{100bp}%
        \@p@@sbburx{200bp}%
        \@p@@sbbury{200bp}%
        \psdraft
      \fi
    \else                       
      \closein1%
      \edef\@p@sfile{\BaseName.bb}%
      \typeout{using BB from \@p@sfile}%
      \ifnum\fig@driver=3
        \edef\@p@sfilefinal{\BaseName.pcx}%
      \else
        \ifepsfig@dos
          \edef\@p@sfilefinal{"`gunzip -c `texfind \BaseName.{z,Z,gz}"}%
        \else
          \edef\@p@sfilefinal{"`epsfig \if@compress-c \fi#1"}%
        \fi
      \fi
    \fi
  \else\closein1                
    \edef\@p@sfile{#1}%
    \if@compress  
      \edef\@p@sfilefinal{"`epsfig -c #1"}%
    \else
      \edef\@p@sfilefinal{#1}%
    \fi
  \fi%
}

\let\@p@@sfigure\@p@@sfile
\def\@p@@sbbllx#1{%
                                            \@bbllxtrue
                \ps@dimcent=#1
                \edef\@p@sbbllx{\number\ps@dimcent}%
                \divide\ps@dimcent by65536
                \global\edef\epsfllx{\number\ps@dimcent}%
}%
\def\@p@@sbblly#1{%
                \@bbllytrue
                \ps@dimcent=#1
                \edef\@p@sbblly{\number\ps@dimcent}%
                \divide\ps@dimcent by65536
                \global\edef\epsflly{\number\ps@dimcent}%
}%
\def\@p@@sbburx#1{%
                \@bburxtrue
                \ps@dimcent=#1
                \edef\@p@sbburx{\number\ps@dimcent}%
                \divide\ps@dimcent by65536
                \global\edef\epsfurx{\number\ps@dimcent}%
}%
\def\@p@@sbbury#1{%
                \@bburytrue
                \ps@dimcent=#1
                \edef\@p@sbbury{\number\ps@dimcent}%
                \divide\ps@dimcent by65536
                \global\edef\epsfury{\number\ps@dimcent}%
}%
\def\@p@@sheight#1{%
                \@heighttrue
                \global\epsfysize=#1
                \ps@dimcent=#1
                \edef\@p@sheight{\number\ps@dimcent}%
}%
\def\@p@@swidth#1{%
                \@widthtrue
                \global\epsfxsize=#1
                \ps@dimcent=#1
                \edef\@p@swidth{\number\ps@dimcent}%
}%
\def\@p@@srheight#1{%
                \@rheighttrue\use@psfigtrue
                \ps@dimcent=#1
                \edef\@p@srheight{\number\ps@dimcent}%
}%
\def\@p@@srwidth#1{%
                \@rwidthtrue\use@psfigtrue
                \ps@dimcent=#1
                \edef\@p@srwidth{\number\ps@dimcent}%
}%
\def\@p@@sangle#1{%
                \use@psfigtrue
                \@angletrue
                \edef\@p@sangle{#1}%
}%
\def\@p@@ssilent#1{%
                \@verbosefalse
}%
\def\@p@@snoisy#1{%
                \@verbosetrue
}%
\def\@cs@name#1{\csname #1\endcsname}%
\def\@setparms#1=#2,{\@cs@name{@p@@s#1}{#2}}%
\def\ps@init@parms{%
                \@bbllxfalse \@bbllyfalse
                \@bburxfalse \@bburyfalse
                \@heightfalse \@widthfalse
                \@rheightfalse \@rwidthfalse
                \def\@p@sbbllx{}\def\@p@sbblly{}%
                \def\@p@sbburx{}\def\@p@sbbury{}%
                \def\@p@sheight{}\def\@p@swidth{}%
                \def\@p@srheight{}\def\@p@srwidth{}%
                \def\@p@sangle{0}%
                \def\@p@sfile{}%
                \use@psfigfalse
                \@prologfilefalse
                \def\@sc{}%
                \if@noisy
                        \@verbosetrue
                \else
                        \@verbosefalse
                \fi
                \@clipfalse
}%
\def\parse@ps@parms#1{%
                \@psdo\@psfiga:=#1\do
                   {\expandafter\@setparms\@psfiga,}%
\if@prologfile
\fi
}%
\def\bb@missing{%
        \if@verbose
            \typeout{psfig: searching \@p@sfile \space  for bounding box}%
        \fi
        \epsfgetbb{\@p@sfile}%
        \ifepsfbbfound
            \ps@dimcent=\epsfllx bp\edef\@p@sbbllx{\number\ps@dimcent}%
            \ps@dimcent=\epsflly bp\edef\@p@sbblly{\number\ps@dimcent}%
            \ps@dimcent=\epsfurx bp\edef\@p@sbburx{\number\ps@dimcent}%
            \ps@dimcent=\epsfury bp\edef\@p@sbbury{\number\ps@dimcent}%
        \else
            \epsfbbfoundfalse
        \fi
}
\newdimen\p@intvaluex
\newdimen\p@intvaluey
\def\rotate@#1#2{{\dimen0=#1 sp\dimen1=#2 sp
                  \global\p@intvaluex=\cosine\dimen0
                  \dimen3=\sine\dimen1
                  \global\advance\p@intvaluex by -\dimen3
                  \global\p@intvaluey=\sine\dimen0
                  \dimen3=\cosine\dimen1
                  \global\advance\p@intvaluey by \dimen3
                  }}%
\def\compute@bb{%
                \epsfbbfoundfalse
                \if@bbllx\epsfbbfoundtrue\fi
                \if@bblly\epsfbbfoundtrue\fi
                \if@bburx\epsfbbfoundtrue\fi
                \if@bbury\epsfbbfoundtrue\fi
                \ifepsfbbfound\else\bb@missing\fi
                \ifepsfbbfound\else
                \@latexerr{ERROR. cannot locate BoundingBox}\@whattodobb
                        \@p@@sbbllx{100bp}%
                        \@p@@sbblly{100bp}%
                        \@p@@sbburx{200bp}%
                        \@p@@sbbury{200bp}%
                        \no@bbtrue
                        \psdraft
                \fi
                \count203=\@p@sbburx
                \count204=\@p@sbbury
                \advance\count203 by -\@p@sbbllx
                \advance\count204 by -\@p@sbblly
                \edef\ps@bbw{\number\count203}%
                \edef\ps@bbh{\number\count204}%
                 \edef\@bbw{\number\count203}%
                \edef\@bbh{\number\count204}%
               \if@angle
                        \Sine{\@p@sangle}\Cosine{\@p@sangle}%
 
{\ps@dimcent=\maxdimen\xdef\r@p@sbbllx{\number\ps@dimcent}%
 
\xdef\r@p@sbblly{\number\ps@dimcent}%
 
\xdef\r@p@sbburx{-\number\ps@dimcent}%
 
\xdef\r@p@sbbury{-\number\ps@dimcent}}%
                        \def\minmaxtest{%
                           \ifnum\number\p@intvaluex<\r@p@sbbllx
                              \xdef\r@p@sbbllx{\number\p@intvaluex}\fi
                           \ifnum\number\p@intvaluex>\r@p@sbburx
                              \xdef\r@p@sbburx{\number\p@intvaluex}\fi
                           \ifnum\number\p@intvaluey<\r@p@sbblly
                              \xdef\r@p@sbblly{\number\p@intvaluey}\fi
                           \ifnum\number\p@intvaluey>\r@p@sbbury
                              \xdef\r@p@sbbury{\number\p@intvaluey}\fi
                           }%
                        \rotate@{\@p@sbbllx}{\@p@sbblly}%
                        \minmaxtest
                        \rotate@{\@p@sbbllx}{\@p@sbbury}%
                        \minmaxtest
                        \rotate@{\@p@sbburx}{\@p@sbblly}%
                        \minmaxtest
                        \rotate@{\@p@sbburx}{\@p@sbbury}%
                        \minmaxtest
 
\edef\@p@sbbllx{\r@p@sbbllx}\edef\@p@sbblly{\r@p@sbblly}%
 
\edef\@p@sbburx{\r@p@sbburx}\edef\@p@sbbury{\r@p@sbbury}%
                \fi
                \count203=\@p@sbburx
                \count204=\@p@sbbury
                \advance\count203 by -\@p@sbbllx
                \advance\count204 by -\@p@sbblly
                \edef\@bbw{\number\count203}%
                \edef\@bbh{\number\count204}%
}%
\def\in@hundreds#1#2#3{\count240=#2 \count241=#3
                     \count100=\count240        
                     \divide\count100 by \count241
                     \count101=\count100
                     \multiply\count101 by \count241
                     \advance\count240 by -\count101
                     \multiply\count240 by 10
                     \count101=\count240        
                     \divide\count101 by \count241
                     \count102=\count101
                     \multiply\count102 by \count241
                     \advance\count240 by -\count102
                     \multiply\count240 by 10
                     \count102=\count240        
                     \divide\count102 by \count241
                     \count200=#1\count205=0
                     \count201=\count200
                        \multiply\count201 by \count100
                        \advance\count205 by \count201
                     \count201=\count200
                        \divide\count201 by 10
                        \multiply\count201 by \count101
                        \advance\count205 by \count201
                     \count201=\count200
                        \divide\count201 by 100
                        \multiply\count201 by \count102
                        \advance\count205 by \count201
                     \edef\@result{\number\count205}%
}%
\def\compute@wfromh{%
                \in@hundreds{\@p@sheight}{\@bbw}{\@bbh}%
                \edef\@p@swidth{\@result}%
}%
\def\compute@hfromw{%
                \in@hundreds{\@p@swidth}{\@bbh}{\@bbw}%
                \edef\@p@sheight{\@result}%
}%
\def\compute@handw{%
                \if@height
                        \if@width
                        \else
                                \compute@wfromh
                        \fi
                \else
                        \if@width
                                \compute@hfromw
                        \else
                                \edef\@p@sheight{\@bbh}%
                                \edef\@p@swidth{\@bbw}%
                        \fi
                \fi
}%
\def\compute@resv{%
                \if@rheight \else \edef\@p@srheight{\@p@sheight} \fi
                \if@rwidth \else \edef\@p@srwidth{\@p@swidth} \fi
}%
\def\compute@sizes{%
        \if@scalefirst\if@angle
        \if@width
           \in@hundreds{\@p@swidth}{\@bbw}{\ps@bbw}%
           \edef\@p@swidth{\@result}%
        \fi
        \if@height
           \in@hundreds{\@p@sheight}{\@bbh}{\ps@bbh}%
           \edef\@p@sheight{\@result}%
        \fi
        \fi\fi
        \compute@handw
        \compute@resv
}
\long\def\graphic@verb#1{\def\next{#1}%
  {\expandafter\graphic@strip\meaning\next}}
\def\graphic@strip#1>{}
\def\graphic@zapspace#1{%
  #1\ifx\graphic@zapspace#1\graphic@zapspace%
  \else\expandafter\graphic@zapspace%
  \fi}
\def\psfig#1{%
\edef\@tempa{\graphic@zapspace#1{}}%
\ifvmode\leavevmode\fi\vbox {%
        \ps@init@parms
        \parse@ps@parms{\@tempa}%
        \ifnum\@psdraft=1
                \typeout{[\@p@sfilefinal]}%
                \if@verbose
                        \typeout{epsfig: using PSFIG macros}%
                \fi
                \psfig@method
        \else
                \epsfig@draft
        \fi
}
}%
\def\graphic@zapspace#1{%
  #1\ifx\graphic@zapspace#1\graphic@zapspace%
  \else\expandafter\graphic@zapspace%
  \fi}
\def\epsfig#1{%
\edef\@tempa{\graphic@zapspace#1{}}%
\ifvmode\leavevmode\fi\vbox {%
        \ps@init@parms
        \parse@ps@parms{\@tempa}%
        \ifnum\@psdraft=1
          \if@angle\use@psfigtrue\fi
          {\ifnum\fig@driver=1\global\use@psfigtrue\fi}%
          {\ifnum\fig@driver=3\global\use@psfigtrue\fi}%
          {\ifnum\fig@driver=4\global\use@psfigtrue\fi}%
          {\ifnum\fig@driver=5\global\use@psfigtrue\fi}%
                \ifuse@psfig
                        \if@verbose
                                \typeout{epsfig: using PSFIG macros}%
                        \fi
                        \psfig@method
                \else
                        \if@verbose
                                \typeout{epsfig: using EPSF macros}%
                        \fi
                        \epsf@method
                \fi
        \else
                \epsfig@draft
        \fi
}%
}%

\def\epsf@method{%
        \epsfbbfoundfalse
        \if@bbllx\epsfbbfoundtrue\fi
        \if@bblly\epsfbbfoundtrue\fi
        \if@bburx\epsfbbfoundtrue\fi
        \if@bbury\epsfbbfoundtrue\fi
        \ifepsfbbfound\else\epsfgetbb{\@p@sfile}\fi
        \ifepsfbbfound
           \typeout{<\@p@sfilefinal>}%
           \epsfig@gofer
        \else
          \@latexerr{ERROR - Cannot locate BoundingBox}\@whattodobb
          \@p@@sbbllx{100bp}%
          \@p@@sbblly{100bp}%
          \@p@@sbburx{200bp}%
          \@p@@sbbury{200bp}%
                \count203=\@p@sbburx
                \count204=\@p@sbbury
                \advance\count203 by -\@p@sbbllx
                \advance\count204 by -\@p@sbblly
                \edef\@bbw{\number\count203}%
                \edef\@bbh{\number\count204}%
          \compute@sizes
          \epsfig@@draft
       \fi
}%
\def\psfig@method{%
        \compute@bb
        \ifepsfbbfound
          \compute@sizes
          \psfig@start
          \vbox to \@p@srheight sp{\hbox to \@p@srwidth 
            sp{\hss}\vss\psfig@end}%
        \else
           \epsfig@draft
        \fi
}%
\def\epsfig@draft{\compute@bb\compute@sizes\epsfig@@draft}%
\def\epsfig@@draft{%
\typeout{<(draft only) \@p@sfilefinal>}%
\if@draftbox
        \hbox{{\fboxsep0pt\fbox{\vbox to \@p@srheight sp{%
        \vss\hbox to \@p@srwidth sp{ \hss 
           \expandafter\Literally\@p@sfilefinal\@nil
                          \hss }\vss
        }}}}%
\else
        \vbox to \@p@srheight sp{%
        \vss\hbox to \@p@srwidth sp{\hss}\vss}%
\fi
}%
\def\Literally#1\@nil{{\tt\graphic@verb{#1}}}
\psfigdriver{dvips}%
\epsfigRestoreAt

\renewcommand{\thefootnote}{\fnsymbol{footnote}}        

\begin{document}

\def\beq{\begin{equation}}
\def\eq{\end{equation}}
\def\eeq{\end{equation}}
\def\bea{\begin{eqnarray}}
\def\ea{\end{eqnarray}}

\newcommand{\newc}{\newcommand}

\newc{\gsim}{\lower.7ex\hbox{$\;\stackrel{\textstyle>}{\sim}\;$}}
\newc{\lsim}{\lower.7ex\hbox{$\;\stackrel{\textstyle<}{\sim}\;$}}

\normalsize\textlineskip

\title{\hfill  \\
\hfill {\normalsize SU-ITP 98-17} \\
\hfill {\normalsize hep-ph/9803420 } \\
\vspace{.3in}
CP-ODD PHASES IN SLEPTON PAIR PRODUCTION\footnote{To appear
in the proceedings of the the Second International Workshop on Electron-Electron
Interactions at TeV Energies.}}


\author{SCOTT THOMAS%
}

\address{Physics Department \\
Stanford University \\ Stanford, CA 94305}



\maketitle\abstracts{
The effects of CP-odd supersymmetric phases on slepton pair production
are considered.
It is shown that CP-even observables
in $e^+ e^-$ and $e^- e^-$ collisions, such as the total selectron cross section,
can depend on CP-odd supersymmetric phases through interference
between different tree level amplitudes.
Left handed selectron pair production in $e^- e^-$ collisions
is particularly sensitive to the
relative phase between the bino and wino masses.
This sensitivity is not limited to any kinematic regime and extends
over all of neutralino parameter space.
The relative phase between the bino and wino masses
is a renormalization group invariant at one-loop,
and as such provides a clean probe for operators which
violate gaugino universality at the messenger scale.
}

\setcounter{footnote}{0}
\renewcommand{\thefootnote}{\alph{footnote}}

\vspace*{1pt}\textlineskip


\section{Introduction}

If nature is supersymmetric at the weak scale, a plethora
of superpartners are waiting to be discovered at future
colliders.
The spectrum and couplings of the superpartners
provide an indirect window to the messenger scale
for supersymmetry breaking.
Precision measurements of the superpartners
could therefore provide indirect information
about physics at scales well beyond those directly
accessible to colliders.
In this paper the possibility of measuring CP-odd
supersymmetric phases
in selectron pair production is considered.



The CP-violating phases of the minimal supersymmetric standard
model (MSSM) are reviewed in the context of slepton pair production
in the next section.
The two basis independent combinations of phases
in the neutralino mass matrix are identified.
The possibility and advantages of measuring CP-odd phases with
 CP-even observables
through interference between different tree level amplitudes
is discussed in section 3.
It is shown that selectron pair production in $e^+ e^-$
or $e^- e^-$ collisions can depend on the CP-odd phases in the
neutralino mass matrix.
These processes are interesting in that effects of CP-odd
phases arise from interference between amplitudes in the
same kinematic channel, and so are not limited to any particular
region of phase space.
The relative sensitivity in different helicity modes to these phases
in the gaugino or Higgsino limit
is explained in section 3.1 in terms of the chiral properties of the
tree level amplitudes.
In section 3.2
the production of left handed sleptons in $e^- e^-$ collisions
is shown to be particularly sensitive to
the relative phase between the bino and wino masses.
This process is unique in that the phase sensitivity extends over all
of supersymmetry parameter space.
Other modes are suppressed outside the mixed gaugino-neutralino
region of parameter space.

The relative phase between the bino and wino masses is a renormalization
group invariant at one-loop, and therefore provides a clean
probe for violations of gaugino universality at the messenger scale.
Precision measurements of left handed slepton production in the
$e^- e^-$ mode of the Next Linear Collider (NLC)
provide an interesting window to the messenger scale through
this phase.
This mode also provides probably the best opportunity to measure
any supersymmetric CP-odd phase at the NLC, and is complimentary
to low energy electric dipole measurements which are
are not directly sensitive to
the relative phase between the bino and wino masses.

\section{CP-Violating Phases in the MSSM}

\label{cpsec}

The CP-violating phases which arise in the MSSM beyond those
of the standard model appear in Lagrangian mass parameters.
The first appears in the superpotential
Dirac mass parameter $\mu$,
\beq
W = \mu H_u H_d
\label{muterm}
\eeq
Assuming squark and slepton
universality, the remaining phases appear in the
soft SUSY breaking mass parameters
$m_i$, $A$, and $m_{ud}^2$,
\beq
{\cal L} =
- {1 \over 2} m_{i} \lambda_i \lambda_i
- A \left(  h_u Q H_u \bar{u}
     - h_d Q H_d \bar{d}
     - h_e L H_d \bar{e} \right)
- m_{ud}^2 H_u H_d
~+~ h.c.
\label{softterms}
\eeq
where $\lambda_i$, $i=1,2,3$ are the gauginos, and $h_i$ are the Yukawa
coupling matrices.
Only a subset of all the phases in the Lagrangian parameters
(\ref{muterm}) and (\ref{softterms}) represent basis
independent combinations of physical CP-violating phases.
The simplest way to determine these basis independent combinations
is to notice that for
$\left\{ \mu, m_i, A, m_{ud}^2 \right\} \to 0$
the MSSM possesses additional $U(1)_{PQ}$ and $U(1)_R$ global
symmetries.\cite{charges}
The mass parameters can therefore be treated as background
spurions which spontaneously break the global symmetries.
A particular assignment of background charges to the mass parameters and fields
is listed in Table 1.
\begin{table}
\tcaption{ Background charges of spurions and fields.
}
\leavevmode
\begin{center}
\label{chargetable}
\begin{tabular}{crr}
\hline
\hline
           & \multicolumn{1}{c}{$U(1)_{PQ}$}
           & \multicolumn{1}{c}{$U(1)_R$}   \\
\hline
$m_i        $ &  0   & $-$2  \\
$A$           &  0   & $-$2  \\
$m_{ud}^2$    & $-$2 &  0  \\
$\mu$         & $-$2 &  2  \\
$H_u$         &    1 &  0  \\
$H_d$         &    1 &  0  \\
$Q \bar{u}$   & $-$1 &  2  \\
$Q \bar{d}$   & $-$1 &  2  \\
$L \bar{e}$   & $-$1 &  2  \\
\hline
\hline
\end{tabular}
\end{center}
\end{table}
%
Physical amplitudes must be invariant under the background symmetries.
The invariant combinations of mass parameters which can
have a non-trivial phase and appear
in physical amplitudes are
%
\beq
A^* m_i  ~~~~ A \mu (m_{ud}^2)^*
\label{Aparameters}
\eq
\beq
m_i \mu ( m_{ud}^2 )^*  ~~~~ m_i^* m_j
\label{mparameters}
\eq
For selectron production 
the gluino
mass does not appear through two loops, and
effects suppressed by the electron Yukawa coupling are irrelevant.
Observable effects of non-zero phases can therefore only appear in
the parameters (\ref{mparameters}) with $i=1,2$.
Among these there are two linear combinations of phases which may
be taken to be
\beq
{\rm Arg}\left(m_1 \mu (m_{ud}^2)^* \right)
 ~~~~~
{\rm Arg} \left( m_1^* m_2 \right )
\label{phases}
\eq
At tree level these phases can affect selectron production only through
the neutralino mass matrix.

In the basis
$( -i \tilde{B}, -i \tilde{W}, \tilde{H}_d, \tilde{H}_u)$
the neutralino mass matrix is
$$
{\cal L} = -{ 1 \over 2} \lambda M \lambda ~+~h.c.
$$
\beq
M = \left(
\begin{array}{cccc}
m_{1}  & 0 & - (g^{\prime} / \sqrt{2} ) H_d^{0*} &
    ( g^{\prime} / \sqrt{2} ) H_u^{0*} \\
0 & m_2 &   (g / \sqrt{2} ) H_d^{0*} &
    - (g / \sqrt{2} ) H_u^{0*} \\
 - (g^{\prime} / \sqrt{2} ) H_d^{0*} &  ({g / \sqrt{2} }) H_d^{0*} &
 0 & -\mu \\
 ({g^{\prime} / \sqrt{2} }) H_u^{0*} &  - ({g / \sqrt{2} }) H_u^{0*} &
 -\mu & 0 \\
\end{array}   \right)
\label{Nmass}
\eq
$$ $$
where $H_u^0$ and $H_d^0$ are understood to be expectation values.
In a general basis all terms in the mass matrix are complex.
The off diagonal gauge interaction
terms which mix the gauginos and Higgsinos clearly
depend on the phase of the Higgs condensates.
The relative phase of the two Higgs condensates is not arbitrary
but determined dynamically by the Higgs potential.
The only tree level potential term which depends
on the relative phase is
\begin{eqnarray}
V &\supset&  - m_{ud}^2 H_u^0 H_d^0 ~+~ h.c.  \nonumber \\
  & & \nonumber \\
  &=& -|m_{ud}^2 H_u^0 H_d^0|  \cos \left[
     {\rm Arg}(m_{ud}^2) + {\rm Arg}(H_u H_d) \right]
\end{eqnarray}
In the ground state, with broken electroweak symmetry,
the phases of the Higgs condensates
dynamically adjust to
${\rm Arg}(m_{ud}^2) = -  {\rm Arg}(H_u H_d)$.
This corresponds to vanishing expectation value for the
pseudo-scalar Higgs boson $A^0$, and is generally not
modified by quantum corrections.
With this alignment, the complex phases which appear in the mass matrix
are those of $m_1, m_2, \mu$, and $m_{ud}^2$.
From the discussion of the basis independent combinations of phases
given above it is clear that a diagonalization of (\ref{Nmass})
can only depend on the two phases (\ref{phases}).
It is these two phases which can have an effect on slepton
pair production as discussed in the next section.
The rotation between mass and interaction eigenstates,
$\lambda = V \chi$, is in general complex for non-zero phases
(\ref{phases}).
It is always possible to work in a basis
in which the mass eigenvalues are real, although it is
sometimes convenient to leave the eigenvalues complex, as
is done in the next section.
It is however not possible in general to absorb all the
phase dependence in (\ref{Nmass}) onto the eigenvalues.

The phase ${\rm Arg}\left(m_1 \mu (m_{ud}^2)^* \right)$
only appears implicitly in the neutralino mass matrix in the
off diagonal mixing terms through the phase
of the Higgs condensates relative to the other parameters.
Its effects are therefore unsuppressed only in the mixed
region of parameter space,
$ \left| |\mu|^2 - |m_1|^2 \right| \lsim m_Z^2$,
in which gaugino-Higgsino mixings are important.
In contrast, the effects of the phase ${\rm Arg} \left( m_1^* m_2 \right )$
do not require mixing and are unsuppressed over all of the
neutralino parameter space.

The phases ${\rm Arg}(A^* m_1)$ and
${\rm Arg}\left(m_1 \mu (m_{ud}^2)^* \right)$ are bounded
by electric dipole moment measurements.
Over most of the SUSY parameter space these phases are typically bounded
to be less than $10^{-(1-3)}$.
Electric dipole measurements are however
not directly sensitive to
${\rm Arg} \left( m_1^* m_2 \right )$.
Mixing effects in the neutralino mass matrix do allow for
non-vanishing contributions to electric dipole moments,
but these are suppressed in the gaugino or Higgsino
limit.
This is in contrast to left handed slepton pair production in
$e^-e^-$ collisions discussed in section 3.2, which has an
unsuppressed sensitivity to ${\rm Arg} \left( m_1^* m_2 \right )$
in these regions of parameter space.
Measurements in this mode of slepton pair production are
therefore complimentary to electric dipole moment measurements.

The expectations for the magnitude of these CP-violating
phases of course depends on the model for the messenger and
supersymmetry breaking sectors.
Under the assumption of gaugino universality, such as would
occur in dilaton dominated supersymmetry breaking,
${\rm Arg} \left( m_1^* m_2 \right )$ would be expected to vanish.
However, even in this class of theories
violations of universality can in general induce
non-zero ${\rm Arg} \left( m_1^* m_2 \right )$.
For example, in the dilaton dominated ansatz, Planck scale slop
can induce a small relative phase between the bino and wino
masses.\cite{nima}
Since the gaugino mass renormalization group equations are
homogeneous at one-loop, the relative phases are preserved
at this order under renormalization group evolution.
Two loop renormalization group
modifications of ${\rm Arg} \left( m_1^* m_2 \right )$
from mixing with ${\rm Arg}(A^* m_i)$ and
${\rm Arg}(m_i^* m_3)$
typically amount to only a fraction of a percent
even for high scale supersymmetry breaking.
The sensitivity of slepton pair production to
${\rm Arg} \left( m_1^* m_2 \right )$ discussed below therefore provides
a clean probe for operators which violate gaugino universality
at the messenger scale.

In the CP-conserving case the phases (\ref{phases}) reduce to two
sign ambiguities in the neutralino mass matrix
\beq
{\rm sgn}( m_1 \mu m_{ud}^2 ) = \pm
 ~~~~~
{\rm sgn}( m_1 m_2) = \pm
\label{signs}
\eq
The first of these is often referred to in the literature
as ${\rm sgn}(\mu)$ with
some particular choice of basis.
The second sign ambiguity is often ignored, and $m_1$ and $m_2$
are tacitly assumed to have the same sign.
Even in the CP-conserving case, the signs
(\ref{signs}) can have very large effects on selectron pair production
as discussed below.






\section{Slepton Pair Production}

Any physical process in general receives contributions from
multiple quantum amplitudes.
The probabilities then depend on both the magnitudes and
relative phases of the amplitudes.
Physical observables can depend on CP-odd phases
directly through the interference of
relative phases of the amplitudes.
This is true {\it even} for CP-even observables.
In this case, since a CP-odd phase
changes sign under CP or T, $\varphi \to - \varphi$,
the observable must depend on the CP or T even quantity
$\cos \varphi$.
Because of this, a CP-even measurements can only determine
a CP-odd phase up to a $Z_2$
ambiguity.\fnm{$\dagger$}\fnt{$\dagger$}{This $Z_2$ sign ambiguity
differs from the sign ambiguities of the
neutralino mass matrix in the CP-conserving case.
The latter sign ambiguities can be determined by CP-even
measurements as discussed below.}
If near destructive interference occurs between two
amplitudes for some values of CP-odd phases, then certain CP-even
observables can in fact be quite sensitive to these phase.
This is in fact the case for left handed selectron production
in $e^- e^-$ collisions as discussed in section 
3.2.

The general
scheme for determining CP-odd phases 
outlined here differs significantly from the standard treatment.
Almost all discussions of the effects of CP-violating phases
in the literature
rely on the fact that amplitudes which depend on CP-odd phases
are conjugated under CP or T.
However, CP-conserving final state rescaterings give an imaginary
contribution to the amplitude which does not change sign under CP or T.
A CP-odd observable then receives a contribution from interference
between the CP-violating and final state amplitudes, proportional
to $\sin \varphi$.
While such CP-odd observables provide a direct probe for CP-violation,
in the absence finite width enhancements for nearly degenerate states,
they are generally unobservably small for supersymmetric phenomena
at colliders since final state rescatterings only occur at one-loop.
In contrast, the CP-even observables described below are
sensitive to CP-odd phases at tree level.

Specializing to the case of charged slepton pair production, it is interesting
to determine which channels are sensitive to supersymmetric CP-violating
phases through interference between different amplitudes.
As discussed in the previous section, since Yukawa coupling
effects are generally irrelevant to production, the only
possible phase dependence arises in the neutralino mass matrix.
Slepton production at hadron colliders proceeds through
s-channel $\gamma^*$ and $Z^{*}$ exchange, and so is not
sensitive to neutralino phases.
The same applies to 
smuon and stau final states at $e^+ e^-$ colliders.
Selectron final states at $e^+ e^-$ and $e^- e^-$ colliders
do in general, however, have contributions from
$t$- and $u$-channel neutralino amplitudes.

\newpage

\subsection{The Neutralino Functions}

In order to discuss selectron production at $e^+ e^-$ and
$e^- e^-$ colliders it is useful to work in the helicity
or equivalently chiral basis.
In this basis the right chiral initial states couple only
to neutralinos through the bino component,
while left chiral initial states couple through both
bino and wino components.
The $t$- and $u$- channel neutralino propagators with these
chiral couplings can be written
in compact form in terms of the neutralino functions introduced
by Peskin.\cite{functions,peskin}
The couplings of the $i$-th neutralino mass eigenstate, $\chi_i$,
to left and right handed chiral states are
\bea
\sqrt{2} e V_{Ri} &=& \sqrt{2} e \left(
    {1 \over \cos \theta_w} V_{1i}  \right) \label{Rcoup}
       \\ & & \nonumber \\
\sqrt{2} e V_{Li} &=& \sqrt{2} e \left(
      { 1 \over 2 \cos \theta_w } V_{1i}
        + { 1 \over 2 \sin \theta_w} V_{2i}  \right)
                \label{Lcoup} \\ \nonumber
\ea
where $\lambda = V \chi$,
and the diagonal mass matrix is $V^t M V$.
The neutralino functions are then defined to be proportional
to the sum over mass eigenstates of the
neutralino propagators
weighted by the chiral couplings (\ref{Rcoup}) and (\ref{Lcoup})
\beq
{\cal N}_{ab}(t) = \sum_i V_{ai}^* {1 \over |m_i|^2 -t} V_{bi}
\eq
\beq
{\cal M}_{ab}(t) = \sum_i V_{ai}^* {m_i \over |m_i|^2 -t} V_{bi}^*
\eq
for $a,b=L,R$.
The functions ${\cal N}_{ab}(t)$ arise from chirally conserving
neutralino propagators, while ${\cal M}_{ab}(t)$ are from the
chirally violating propagators.
The contributions of the four mass eigenstates to the neutralino
functions in general have some non-trivial interference.
Peskin's dimensionless neutralino functions,
$N_{ab}(t)$ and $M_{ab}(t)$,
are
related to these by
$N_{ab}(t) = |m_1|^2 {\cal N}_{ab}(t)$ and
$M_{ab}(t) = |m_1| {\cal M}_{ab}(t)$.
%

\begin{table}
\tcaption{Selectron production modes which are sensitive to neutralino
phases through interference. Summary of the chiral structure of the
neutralino propagator,
overall magnitude of the cross section, and sensitivity
of the cross section to neutralino phases
in the gaugino or Higgsino limit.
}
\leavevmode
\begin{center}
\label{summary}
\begin{tabular}{cccc}
\hline
\hline
        Mode & Neutralino Propagator &
               \multicolumn{2}{c}{Gaugino/Higgsino Limit} \\
                        &     &    Magnitude  & Phase Sensitivity \\
\hline
$e_R^+ e_R^- \to \tilde{e}_L^+ \tilde{e}_R^-$  &
     Chirally Violating & Suppressed  & Unsuppressed \\
$e_L^+ e_L^- \to \tilde{e}_R^+ \tilde{e}_L^-$  &
     Chirally Violating & Suppressed  & Unsuppressed \\
$e_R^- e_R^- \to \tilde{e}_R^- \tilde{e}_R^-$  &
     Chirally Violating & Unsuppressed  & Suppressed \\
$e_L^- e_R^- \to \tilde{e}_L^- \tilde{e}_R^-$  &
     Chirally Conserving & Suppressed  & Unsuppressed \\
$e_L^- e_L^- \to \tilde{e}_L^- \tilde{e}_L^-$  &
     Chirally Violating & Unsuppressed  & Unsuppressed \\
\hline
\hline
\end{tabular}
\end{center}
\end{table}
%

For non-zero phases (\ref{phases}) in the neutralino mass matrix,
the chirally violating propagator functions ${\cal M}_{ab}(t)$ are
complex in general.
The differential cross section for selectron production with
$e^+ e^-$ and $e^- e^-$
initial states with a net chirality therefore in general depends on the
neutralino phases via interference between the neutralino mass eigenstates.
The chirally conserving propagator function ${\cal N}_{LR}(t)$
also has non-trivial interference in general.
However Im${\cal N}_{aa}(t)=0$ as a result of hermiticity of
the chirally conserving bino-bino and wino-wino propagators.
This has the consequence that for $e^+ e^-$ collisions, modes with pairs of
right handed selectrons or pairs of left handed selectrons in the
final state do not depend on the phases
(\ref{phases}) through interference between different amplitudes.
The remaining modes which are sensitive to neutralino phases through
interference are listed in Table 2.
The differential cross sections for these modes are
\bea
{ d \sigma \over dt}
  (e^+_R e^-_R \rightarrow \tilde{e}^+_L \tilde{e}^-_R) &=&
  3 R ~ \left| {\cal M}_{LR}(t) \right|^2 \\  & &  \nonumber \\
{ d \sigma \over dt}
  (e^+_L e^-_L \rightarrow \tilde{e}^+_R \tilde{e}^-_L) &=&
  3 R ~ \left| {\cal M}_{LR}(t) \right|^2  \\ & & \nonumber \\
{ d \sigma \over dt}
  (e^-_R e^-_R \rightarrow \tilde{e}^-_R \tilde{e}^-_R) &=&
{3 \over 2} R ~ \left| {\cal M}_{RR}(t) +  {\cal M}_{RR}(u)
  \right|^2 \\ & & \nonumber \\
{ d \sigma \over dt}
  (e^-_L e^-_L \rightarrow \tilde{e}^-_L \tilde{e}^-_L) &=&
{3 \over 2} R ~ \left| {\cal M}_{LL}(t) +  {\cal M}_{LL}(u)
  \right|^2  \\ & & \nonumber \\
{ d \sigma \over dt}
  (e^-_L e^-_R \rightarrow \tilde{e}^-_L \tilde{e}^-_R) &=&
3 R \left[ { (t-m_{\tilde{e}_L}^2) (m_{\tilde{e}_R}^2 -t) \over s}
         -t   \right]
~ \left| {\cal N}_{LR}(t)
  \right|^2 \\ 
  \nonumber
\ea
where $R = \sigma(e^+ e^- \rightarrow \mu^+ \mu^-)
= 4 \pi \alpha^2 / 3s$, and the angular
integrations are over $-1 \leq \cos \theta \leq 1$.
The first four of these are s-wave near threshold, while
the last one is p-wave.

An important feature of the
modes listed above is that the phase dependent interference
takes place between different neutralino mass eigenstates
in the {\it same} channel.
Because the interference is between amplitudes in the same
kinematic channel, the effects are not limited to a particular
kinematic region of phase space, and are not suppressed
for production well above threshold.
This is in contrast to analogous chargino and neutralino
processes mentioned in the conclusions.

The magnitudes and phases of the neutralino eigenstate
contributions to the neutralino functions are determined
by diagonalization of the neutralino mass matrix (\ref{Nmass}).
In order to understand the physical content of this
diagonalization it is instructive to consider the mostly gaugino
or mostly Higgsino limit.
In this limit the physical mass eigenstates are mostly
the gaugino and Higgsino eigenstates with small admixtures
of the other states induced by the off diagonal mixing terms
in (\ref{Nmass}).
This limit is reached if the level splitting between the
gauginos and Higgsinos is large compared with the off diagonal
mixing terms,
$ \left| |\mu|^2 - |m_1|^2 \right| \gsim m_Z^2$.
This limit holds regardless of whether the lightest neutralino
is gaugino or Higgsino like, subject to the small mixing
criterion above.

In the gaugino or Higgsino limit the 
bino-wino propagators only arise through mixing
with intermediate Higgsino states.
The bino-wino propagators projected onto the physical mass
eigenstates are therefore
suppressed by ${\cal O}( m_Z^2 / ( \mu^2 - m_i^2) )$
as compared with the bino-bino or wino-wino propagators.
This has the effect that all
the $LR$ neutralino functions are suppressed in magnitude
by a similar amount compared with the $RR$ or $LL$ functions.
These functions, although reduced in magnitude,
are sensitive to the phases in the neutralino mass matrix.

The chirally violating neutralino function
${\cal M}_{LL}(t)$ receives contributions
from both bino-bino and wino-wino
propagators.
In the gaugino or Higgsino limit these propagators projected
onto the physical mass eigenstates are dominated by the mostly
bino and mostly wino states.
Interference between these two amplitudes is therefore
sensitive in this limit to the relative phase between the bino
and wino masses,
${\rm Arg}(m_1^* m_2)$.
However, the chirally violating neutralino function
${\cal N}_{RR}(t)$
is dominated only the mostly bino state
and can not have a large interference with the other states
in this limit.
So while the magnitude of this function is
not suppressed in this limit, its sensitivity to phases is
suppressed.

The (non)suppressions of the overall rate and phase sensitivity in the
gaugino or Higgsino limit for the
$e^+e^-$ and $e^-e^-$ modes discussed above are summarized in Table 2.
The relative suppression of some of the phase sensitive
modes is best illustrated
by considering the pure gaugino or Higgsino limit, in which
case the relevant neutralino functions reduce to
\bea
{\cal M}_{RR}(t) &=&  {1 \over \cos^2 \theta_w}
  {m_1 \over |m_1|^2 -t} \\ & & \nonumber \\
{\cal M}_{LL}(t) &=& {1 \over 4 \sin^2 \theta_w}
 \left( { m_1 \tan^2 \theta_w \over |m_1|^2 -t} +
 { m_2 \over |m_2|^2 -t} \right ) \\ & & \nonumber \\
{\cal M}_{LR}(t)&=&0 \\ & & \nonumber \\
{\cal N}_{LR}(t)&=&0 \\ \nonumber
\ea
Since there is no bino-wino mixing in this limit,
the $LR$ functions vanish.
The chirally violating function ${\cal M}_{LL}(t)$ is given
by pure bino and wino exchange in this limit, and as such the interference
term in $|{\cal M}_{LL}(t)|^2$ depends on
${\rm Arg}(m_1^* m_2)$.
The chirally violating function ${\cal M}_{LL}(t)$ is given by
pure bino exchange in this limit, and
$|{\cal M}_{RR}(t)|^2$ therefore does not depend on either phase in the
neutralino mass matrix.

In the mixed region of neutralino parameter space the differential
cross sections for all the modes
listed in Table 2 depend in a non-trivial way
on the phases (\ref{phases}).
However, it is important to note that even in the gaugino
or Higgsino limit the suppressed modes still have non-trivial
dependence on the phases.
It is in fact these suppressed modes which would help to
determine the $\mu$ parameter in the gaugino limit if the
heavier neutralino states are kinematically inaccessible.
This makes clear that any precision fit of neutralino parameters
to data must include the phases (\ref{phases}).
Even if CP-conservation is assumed the sign ambiguities (\ref{signs})
must be included a fit.

\subsection{Phase Dependence in $e^-e^-$ Collisions}

\label{eesec}

The only mode of selectron pair production in
which both the rate and phase sensitivities are unsuppressed
in the gaugino or Higgsino limit
is $e_L^- e_L^- \to \tilde{e}_L^- \tilde{e}_L^-$.
It is therefore worthwhile to consider this process in the
pure gaugino or Higgsino limit.
The magnitude squared of the neutralino function for this
case is
$$
\left| {\cal M}_{LL}(t) \right|^2=
  {1 \over 16 \sin^4 \theta_w} \left(
{ |m_1|^2 \tan^4 \theta_w \over (|m_1|^2-t)^2 } +
{ 2 |m_1||m_2| \tan^2 \theta_w \cos \left( {\rm Arg}(m_1^*m_2) \right)
     \over (|m_1|^2-t) (|m_2|^2-t) }   \right.
$$
\beq
 \left.
 + { |m_2|^2 \over (|m_2|^2-t)^2 }
 \right)  ~~~~~~~~~\\ \nonumber
\eq
The bino-wino interference term depends on
$\cos \left( {\rm Arg}(m_1^*m_2) \right)$.
Note that maximal constructive(destructive) interference
is obtained for ${\rm Arg}(m_1^* m_2)=0,\pi$
or in the CP-conserving case
${\rm sgn}(m_1 m_2) = +(-)$.
\begin{figure}[htbp]
\centerline{
\epsfig{figure=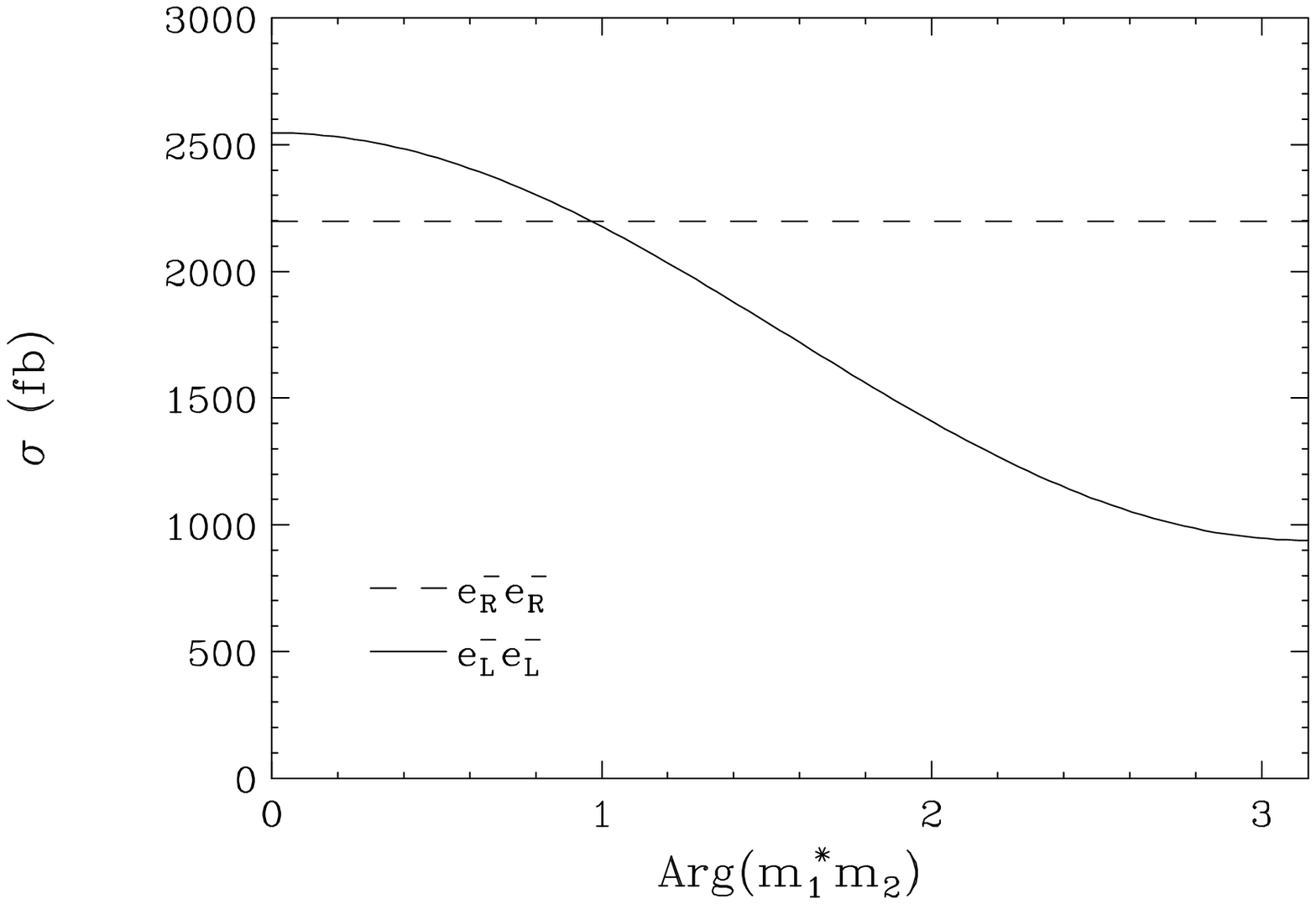,height=6.9cm,width=10.0cm,angle=0}}
\fcaption{Total cross sections for
$e_R^- e_R^- \to \tilde{e}_R^- \tilde{e}_R^-$  and
$e_L^- e_L^- \to \tilde{e}_L^- \tilde{e}_L^-$ in the
pure gaugino or Higgsino limit as a function of ${\rm Arg}(m_1^* m_2)$.
The parameters are
$\sqrt{s} = 500$ GeV, $|m_1| = 100$ GeV, $|m_2|= 200$ GeV,
$m_{\tilde{e}_R} = 130$ GeV, and
$m_{\tilde{e}_L} = 180$ GeV.
For reference $R \simeq 400$ fb at this center of mass energy.
}
\label{totalsig}
\end{figure}
The total cross sections
$\sigma(e_L^- e_L^- \to \tilde{e}_L^- \tilde{e}_L^-)$ and
$\sigma(e_R^- e_R^- \to \tilde{e}_R^- \tilde{e}_R^-)$
are shown in Fig. \ref{totalsig}~as a function of ${\rm Arg}(m_1^* m_2)$
with a typical set of parameters at the NLC.
The differential cross sections for the same set of parameters
are shown in Fig. \ref{difsiga}.
\begin{figure}[htbp]
\centerline{
\epsfig{figure=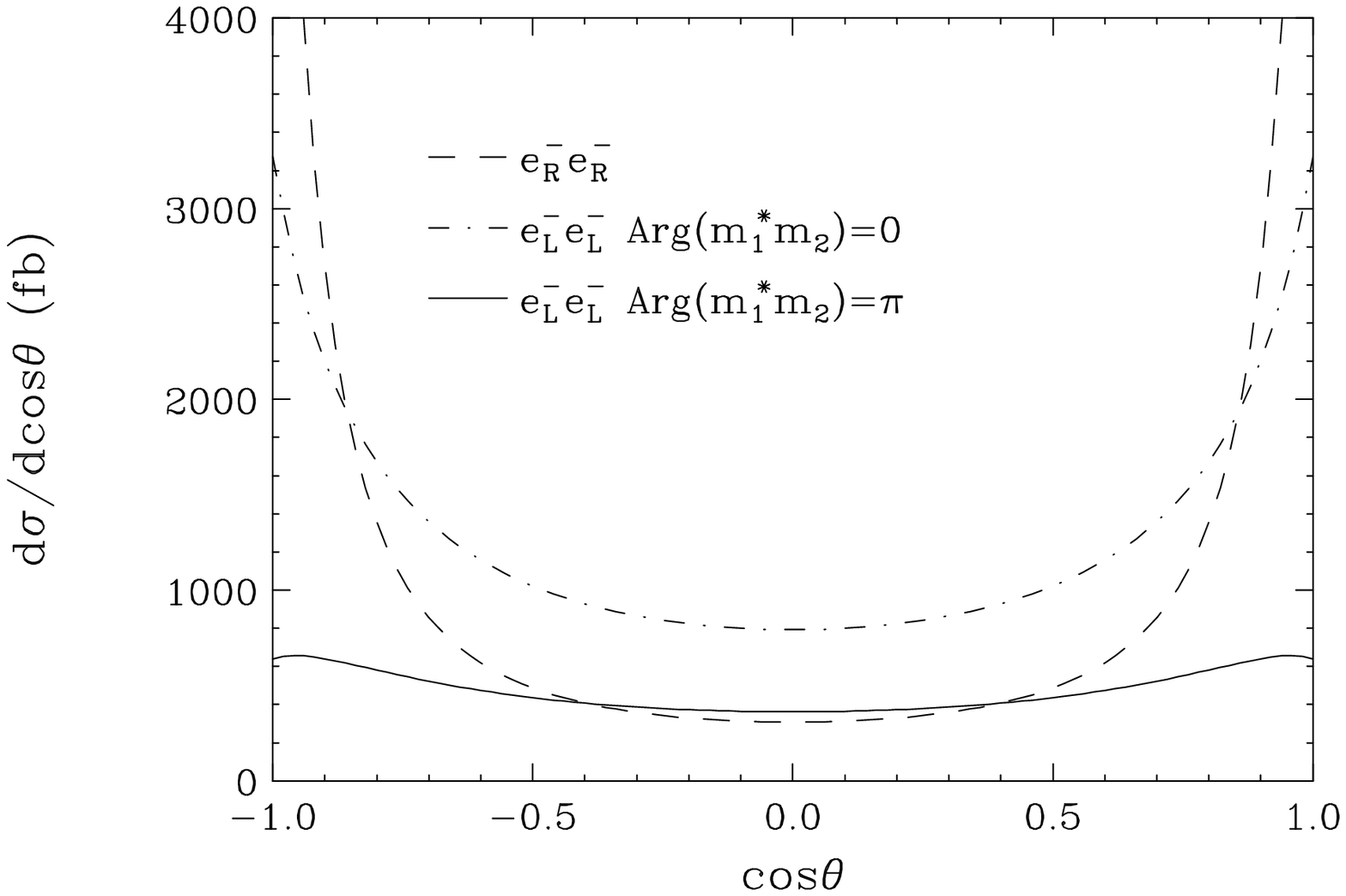,height=6.9cm,width=10.4cm,angle=0}}
\fcaption{Differential cross sections for
$e_R^- e_R^- \to \tilde{e}_R^- \tilde{e}_R^-$  and
$e_L^- e_L^- \to \tilde{e}_L^- \tilde{e}_L^-$ in the
pure gaugino or Higgsino limit for ${\rm Arg}(m_1^* m_2)=0,\pi$.
The parameters are the same those in fig. \ref{totalsig}.
}
\label{difsiga}
\end{figure}
For constructive interference between bino and wino exchange
in $e_L^- e_L^- \to \tilde{e}_L^- \tilde{e}_L^-$
the differential cross section is a monotonic
function of $\cos \theta$ in the forward hemisphere.
Any non-monotonic deviation implies
${\rm Arg}(m_1^*m_2) \neq 0$.
The sensitivity to ${\rm Arg}(m_1^*m_2)$
is very pronounced in the forward direction.

\begin{figure}[htbp]
\centerline{
\epsfig{figure=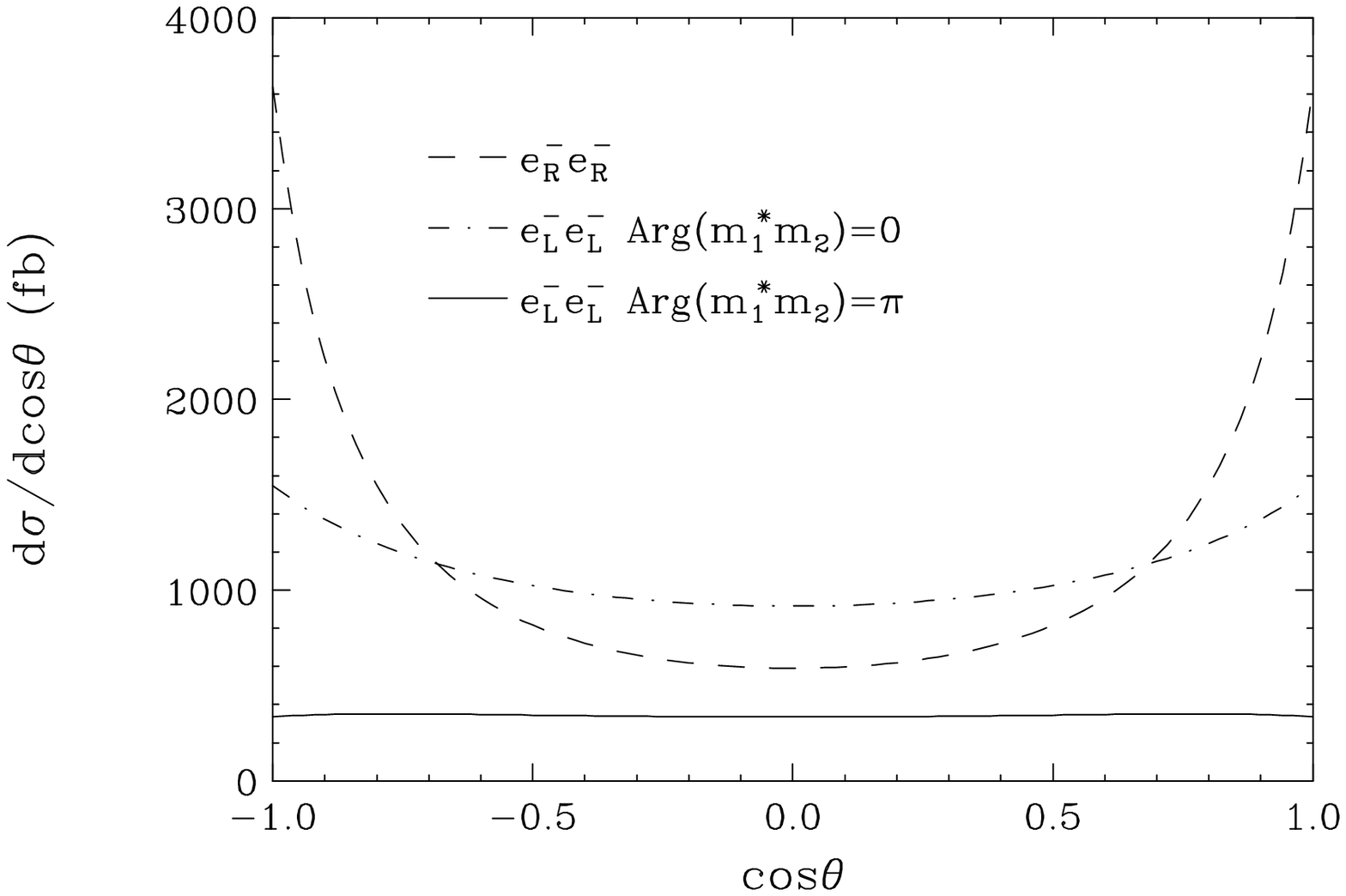,height=6.9cm,width=10.4cm,angle=0}}
\fcaption{Differential cross sections for
$e_R^- e_R^- \to \tilde{e}_R^- \tilde{e}_R^-$  and
$e_L^- e_L^- \to \tilde{e}_L^- \tilde{e}_L^-$ in the
pure gaugino or Higgsino limit for ${\rm Arg}(m_1^* m_2)=0,\pi$.
The parameters are
$\sqrt{s} = 500$ GeV, $|m_1| = 150$ GeV, $|m_2|= 300$ GeV,
$m_{\tilde{e}_R} = 170$ GeV, and
$m_{\tilde{e}_L} = 210$ GeV.}
\label{difsigb}
\end{figure}

The large forward peak in both
$e_R^- e_R^- \to \tilde{e}_R^- \tilde{e}_R^-$  and
$e_L^- e_L^- \to \tilde{e}_L^- \tilde{e}_L^-$ is less
pronounced for scattering near threshold, and with
heavier neutralinos.
The differential cross sections for slightly more massive
states are shown in Fig. \ref{difsigb}.
For this set of parameters the
$e_L^- e_L^- \to \tilde{e}_L^- \tilde{e}_L^-$ distribution
happens to be nearly flat in $\cos \theta$ for destructive interference.

The process $e_L^- e_L^- \to \tilde{e}_L^- \tilde{e}_L^-$
is apparently quite sensitive to the phase
${\rm Arg}(m_1^*m_2)$.
This fortuitous sensitivity is the result of a numerical coincidence which
allows near complete
constructive or destructive interference over much of the kinematic
phase space.
Well above threshold the total cross section is dominated
by small $t$ corresponding to scattering
in the nearly forward direction.
For $t=0$
\beq
\left| {{\cal M}}_{LL}(0) \right|^2=
  {1 \over 16 \sin^4 \theta_w} \left(
 \tan^4 \theta_w +
 {2 |m_1| \over |m_2|} \tan^2 \theta_w
       \cos\left( {\rm Arg}(m_1^*m_2) \right)  +
{ |m_1|^2 \over |m_2|^2 }
\right)
\eq
The interference between bino and wino exchange can
be nearly completely destructive
for ${\rm Arg}(m_1^*m_2) = \pi$
if $\tan^2 \theta_w \sim |m_1 / m_2|$.
This is in fact the case for the gaugino unification value
of $|m_1 / m_2| \simeq {1 \over 2}$.
In the pure gaugino limit, assuming gaugino unification,
$|{\cal M}_{LL}(0)|^2 \simeq$ 0.82(0.04)
for ${\rm Arg}(m_1^*m_2) = 0(\pi)$.

Away from the pure gaugino or pure Higgsino limit
these results are modified slightly, but the qualitative
features are not changed at lowest order.
The overall rates for $e_R^- e_R^- \to \tilde{e}_R^- \tilde{e}_R^-$  and
$e_L^- e_L^- \to \tilde{e}_L^- \tilde{e}_L^-$ are modified
at ${\cal O}(m_Z^2 / (\mu^2 - m_i^2))$ by mixing with the
Higgsino states.
The phase ${\rm Arg}(m_1^*m_2)$, interpreted as the relative
phase of the masses of the mostly bino and mostly wino
states, is also renormalized at ${\cal O}(m_Z^2 / (\mu^2 - m_i^2))$
by ${\rm Arg}\left(m_1 \mu (m_{ud}^2)^* \right)$ through mixing
with the Higgsino states.
An upper limit on this modification can however be
obtained from the electron electric dipole moment experimental bound.
Finally, the rate for $e_L^- e_R^- \to \tilde{e}_L^- \tilde{e}_R^-$
is down by ${\cal O}(m_Z^4 / (\mu^2 - m_i^2)^2)$
compared with the unsuppressed modes.

A precision measurement of ${\rm Arg}(m_1^*m_2)$ at the NLC
in the $e^- e^-$ mode must contend with uncertainties
in the kinematic masses, beam polarization,
supersymmetry parameters responsible for
mixing with the Higgsino states, and detector monte carlo uncertainties
in both signal and background efficiencies with cuts.
For typical parameters at the NLC it is estimated that
even from a total cross section measurement,
${\rm Arg}(m_1^*m_2)$ could be determined to a precision of
a few percent in a year of running.\cite{big}
A slightly better measurement could be obtained
from observables which are optimized to take account of angular
dependences.
As discussed in section 2 this phase is
a clean probe of gaugino universality violations at
the messenger scale.
So the mode $e_L^- e_L^- \to \tilde{e}_L^- \tilde{e}_L^-$
at the NLC provides an interesting window to the messenger scale
for supersymmetry breaking.


\section{Conclusions}

CP-odd phases in the neutralino mass matrix have an effect
on CP-even observables associated with
selectron production in $e^+ e^-$ and $e^- e^-$ collisions
through interference between different tree level amplitudes
in the same kinematic channel.
In the gaugino or Higgsino limit, left handed slepton
production in $e^- e^-$ collisions is very sensitive to the
relative phase between the bino and wino masses.
Since this phase is not modified at a significant
level by renormalization group evolution, this process
provides a clean probe for gaugino non-universality
at the messenger scale.
This is in contrast to small violations of gaugino
universality in the magnitudes of the gaugino masses 
which can receive significant modifications from two-loop
renormalization group evolution.

In addition to the differential cross section for longitudinally
polarized beams, it is possible to form other
CP-even observables which are sensitive to interference between
different amplitudes
with transversely polarized beams.
These are however suppressed outside the mixed
gaugino-Higgsino regions of parameter space, and
require interference between different kinematic
channels.\cite{transverse}
Other possibilities for observing CP-odd phases
in CP-even observables from interference effects are for
$\chi_i^0 \chi_j^0$, $i \neq j$ and
$\chi_1^{\pm} \chi_2^{\mp}$ final states.
These however are suppressed outside the mixed gaugino-Higgsino
regions of parameter space and require a mass insertion in the final
state so are additionally suppressed well above
threshold.\cite{transverse}
Left handed selectron production in $e^- e^-$
collisions therefore provides probably
the best opportunity to measure a CP-odd phase at the NLC,
and is complimentary to low energy electric dipole measurements.

\nonumsection{Acknowledgements}
\noindent
Estimates of the experimental sensitivity to CP-odd phases at the NLC
were done in collaboration with J. Feng.
I would also like to thank H. Haber and S. Martin for useful discussions,
and C. Heusch and N. Rogers for organizing the
Second International Workshop on Electron-Electron
Interactions at TeV Energies.
This work was partially supported by Stanford University through
a Fredrick E. Terman Fellowship.

\nonumsection{References}

\end{document}